\documentclass[conference]{IEEEtran}
\newtheorem{theorem}{Theorem}
\newtheorem{example}{Example}
\newtheorem{algorithm}{Algorithm}

\ifCLASSINFOpdf
\else
\fi

\begin{document}
%
\title{The bitwise operations related to a fast sorting algorithm}

\author{\IEEEauthorblockN{Krasimir Yordzhev}
\IEEEauthorblockA{Faculty of Mathematics and Natural Sciences\\
South-West University\\
 Blagoevgrad, Bulgaria\\
Email: yordzhev@swu.bg}
}

\maketitle

\begin{abstract}
In the work we discuss the benefit of using bitwise operations in programming. Some interesting examples in this respect have been shown. What is described in detail is an algorithm for sorting an integer array with the substantial use of the bitwise operations. Besides its correctness we strictly prove that the described algorithm works in time O(n). In the work during the realisation of each of the examined algorithms we use the apparatus of the object-oriented programming with the syntax and the semantics of the programming language C++.
\end{abstract}


\IEEEpeerreviewmaketitle

\section{Introduction}

The use of bitwise operations is a powerful means during programming with the languages C/C++ and Java. Some of the strong sides of these programming languages are the possibilities of low level programming. Some of the means for this possibility are the introduced standard bitwise operations, with the help of which it is possible to directly operate with every bit of an arbitrary variable situated in the random access memory of the computer. In the current article we are going to describe some methodical aspects for work with the bitwise operations.

As an interesting example of application of the bitwise operations comes the realised by us algorithm for sorting an integer array, for which we strictly prove its correctness and the fact that this algorithm will use $O(n)$   operations included in the standard of the programming language C++. A main role in the realisation of the algorithm play the bitwise operations.

\section{Bitwise operations}

The bitwise operations can be applied for integer data type only. For the definition of the bitwise operations and some of their elementary applications could be seen, for example, in \cite{Davis,Kernigan} for C/C++ programming languages and in \cite{Flanagan,Schildt} for Java programming language.

We assume, as usual that bits numbering in variables starts from right to left, and that the number of the very right one is 0.

Let \verb"x" and \verb"y" be integer variables or constants and let \verb"z" be integer variables of one type, for which $w$ bits are needed. Let \verb"x" and \verb"y" be initialized (if they are variables) and let the assignment \verb"z = x & y;" (bitwise AND), or \verb"z = x | y;" (bitwise inclusive OR), or \verb"z = x ^ y;" (bitwise exclusive OR), or \verb"z = ~x;" (bitwise NOT) be made. For each $i=0,1,2,\ldots ,w-1$, the new contents of the $i$-th bit in \verb"z" will be as it is presented in the Table \ref{bitwise}.

\begin{table}
\caption{Bitwise operations in programming languages C/C++ and Java}\label{bitwise}
\begin{center}
\begin{tabular}{||c|c||c|c|c|c||}
  \hline\hline
$i$ bit of & $i$ bit of & $i$ bit of & $i$ bit of &  $i$ bit of &  $i$ bit of \\
 x &  y &  z = x \& y; &  z = x $\vert$ y; &  z = x \verb"^" y; & z = \verb"~"x; \\
\hline\hline
  0 & 0 & 0 & 0 & 0 & 1\\ \hline
  0 & 1 & 0 & 1 & 1 & 1\\ \hline
  1 & 0 & 0 & 1 & 1 & 0\\ \hline
  1 & 1 & 1 & 1 & 0 & 0\\ \hline
  \hline
\end{tabular}
\end{center}
\end{table}

In case that $k$ is a nonnegative integer, then the statement \verb"z = x<<k;" (bitwise shift left) will write in the $(i+k)$ bit of \verb"z" the value of the $k$ bit of x, where $i=0,1,\ldots ,w-k-1$, and the very right $k$ bits of \verb"x" will be filled by zeroes. This operation is equivalent to a multiplication of \verb"x" by $2^k$.

The statement \verb"z=x>>k" (bitwise shift right) works the similar way. But we must be careful if we use the programming language C or C++, as in various programming environments this operation has different interpretations - somewhere $k$ bits of \verb"z" from the very left place are compulsory filled by 0 (logical displacement), and elsewhere the very left $k$ bits of \verb"z" are filled with the value from the very left (sign) bit; i.e. if the number is negative, then the filling will be with 1 (arithmetic displacement). Therefore it is recommended to use unsigned type of variables (if the opposite is not necessary) while working with bitwise operations (see also Example \ref{Div2func}). In the Java programming language, this problem is solved by introducing the two different operators: \verb"z=x>>k" and \verb"z=x>>>k" \cite{Flanagan,Schildt}.

Bitwise operations are left associative.

The priority of operations in descending order is as follows: \verb"~" (bitwise NOT); the arithmetic operations \verb"*" (multiply), \verb"/" (divide), \verb"%" (remainder or modulus); the arithmetic operations \verb"+" (addition) \verb"-" (subtraction); the bitwise operations \verb"<<" and \verb">>"; the relational operations \verb"<", \verb">", \verb"<=", \verb">=", \verb"==", \verb"!="; the bitwise operations \verb"&",\verb"^" and \verb"|"; the logical operations \verb"&&" and \verb"||".

\section{Some elementary examples of using the bitwise operations}

\begin{example}
To compute the value of the  \verb"i"-th bit (0 or 1) of an integer variable \verb"x" we can use the function:
\end{example}
\begin{verbatim}
int BitValue(int x, unsigned int i) {
    int b = ((x & 1<<i) == 0) ? 0 : 1;
    return b;
}
\end{verbatim}

\begin{example}
Directly from the definition of the operation bitwise shift left (\verb"<<") follows the efficiency of the following function computing  $2^n$, where $n$ is a nonnegative integer:
\end{example}
\begin{verbatim}
unsigned int Power2(unsigned int n) {
    return 1<<n;
}
\end{verbatim}

\begin{example}\label{Div2func}
The integer function $f(x)=x\; \% \; 2^n$ implemented using operation bitwise shift right (\verb">>").
\end{example}
\begin{verbatim}
int Div2(int x, unsigned int n) {
int s = x<0 ? -1 : 1;
\end{verbatim}
/* s = the sign of x */
\begin{verbatim}
x = x*s;
\end{verbatim}
/*  We reset the sign bit of x */
\begin{verbatim}
return (x>>n)*s;
}
\end{verbatim}

When we work with negative numbers we must consider that in the computer the presentation of the negative numbers is through the so called true complement code. The following function gives us how to code the integers in the memory of the computer we work with. For simplicity we are going to work with type short, but it is not a problem for the function to be overloaded for other integer types, too.

\begin{example}\label{BinReplfunc}
A function showing the presentation of the numbers of type short in the memory of the computer.
\begin{verbatim}
void BinRepl(short n) {
int b;
int d = sizeof(short)*8 - 1;
while (d>=0) {
    b= 1<<d & n ? 1 : 0;
    cout<<b;
    d--;
    }
}
\end{verbatim}
\end{example}

Some experiments with the function \verb"BinRepl" are given in Table \ref{BinReplTable}.

\begin{table}
\begin{center}
\caption{ Presentation of some numbers of type short in the memory of the computer}\label{BinReplTable}
\begin{tabular}{||c||c||}
  \hline\hline
An integer of type  \verb"short" & Presentation in memory  \\
\hline\hline
  0  & 0000000000000000  \\ \hline
  1  & 0000000000000001  \\ \hline
  -1 & 1111111111111111  \\ \hline
  2  & 0000000000000010  \\ \hline
 -2  & 1111111111111110  \\ \hline
 $16=2^4$   & 0000000000010000  \\ \hline
$-16=-2^4$  & 1111111111110000  \\ \hline
 $26=2^4 +2^3 +2$    & 0000000000011010  \\ \hline
$-26=-(2^4 +2^3 +2)$ & 1111111111100110  \\ \hline
 $41=2^5 +2^3 +1$    & 0000000000101001  \\ \hline
$-41=-(2^5 +2^3 +1)$ & 1111111111010111  \\ \hline
 $32767=2^{15} -1$   & 0111111111111111  \\ \hline
$-32767=-(2^{15}-1)$ & 1000000000000001  \\ \hline
$32768=2^{15}$       & 1000000000000000  \\ \hline
$-32768=-2^{15}$     & 1000000000000000  \\ \hline
  \hline
\end{tabular}
\end{center}
\end{table}

Compare the function presented in Example \ref{BinReplfunc} to the next function presented in Example \ref{dectobinfunc}.

\begin{example}\label{dectobinfunc}
A function that prints a given integer in binary notation.
\end{example}
\begin{verbatim}
void DecToBin(int n) {
\end{verbatim}
/*  Prints the sign - , if n<0: */
\begin{verbatim}
    if (n<0) cout<<'-';
    n = abs(n);
    int b;
    int d = sizeof(int)*8 - 1;
\end{verbatim}
/*  Skips the insignificant zeroes at the
    beginning: */
\begin{verbatim}
    while ( d>0 && (n & 1<<d ) == 0 ) d--;
    while (d>=0) {
        b= 1<<d & n ? 1 : 0;
        cout<<b;
        d--;
    }
}
\end{verbatim}

\begin{example}
The following function calculates the number of 1 in a given integer $n$ written in a binary notation. Here again we ignore the sign of the number (if it is negative) and we work with its absolute value.
\begin{verbatim}
int NumbOf_1(int n) {
    n = abs(n);
    int temp=0;
    int d = sizeof(int)*8 - 1;
    for (int i=0; i<d; i++)
        if (n & 1<<i) temp++;
    return temp;
}
\end{verbatim}
\end{example}

\section{Bitwise sorting}

In this section we are going to suggest a fast algorithm for sorting an arbitrary integer array. And since during its realisation we are substantially going to use bitwise operations, we will call it ''\textbf{Bitwise sorting}''. We will prove that the bitwise sorting works in time $O(n)$, where $n$   is the size of the array. This is an excellent evaluation regarding the criterion time. For comparison below we give some of the most famous sorting algorithms and their evaluations by criterion time \cite{Darlington,Knuth,Sedgewick}.

\begin{itemize}
\item	\textbf{Selection sort} – works in time $O(n^2 )$;
\item	\textbf{Bubble sort}– works in time $O(n^2 )$;
\item	\textbf{Bidirectional bubble sort} (\textbf{Cocktail sort}) – works in time $O(n^2 )$;
\item	\textbf{Insertion sort} – works in time $O(n^2 )$;
\item	\textbf{Merge sort} – works in time  $O(n\log n)$;
\item	\textbf{Tree sort} – works in time  $O(n\log n)$;
\item	\textbf{Timsort} – works in time  $O(n\log n)$;
\item	\textbf{Counting sort} – works in time  $O(n+m)$, where $m$  is another parameter, giving the number of the unique keys;
\item	\textbf{Bucket sort} – works in time  $O(n)$.
\end{itemize}

\textbf{Notes}:
\begin{enumerate}
\item	Timsort has been developed for use with the programming language Python \cite{Hetlan}.
\item	For Counting sort  $O(n+m)$  additional memory is necessary.
\item	For Bucket sort $O(m)$  additional memory is necessary, where $m$  is another parameter, giving the number of the unique keys, and it is also necessary to have knowledge of the nature of the sorted data which goes beyond the functions ''swap'' and ''compare''.
\end{enumerate}

\section{Programme code of the algorithm}

The algorithm created by us, described with the help of programming language C++, is shown below (algorithm \ref{biwisesortingalgor}). Due to some obvious reasons, first we create a function which sorts an array whose elements are either only nonnegative, or only negative. The second function divides the given array into two disjoint subarrays respectively only with negative and only with nonnegative elements. After sorting each one of these subarrays, we merge them so that we obtain one finally sorted array.

 \begin{algorithm}\label{biwisesortingalgor}
 Bitwise sorting.
\end{algorithm}

\begin{verbatim}
template <class T>
\end{verbatim}
/* The first function sorts integer elements with
        the same signs:  */
\begin{verbatim}
void BitwiseSort1(T A[],int n){
    int t;
\end{verbatim}
/*  t –- size of the type T in bits  */
\begin{verbatim}
    t = sizeof(T)*8;
    T A0[n], A1[n];
 \end{verbatim}
/*  A0 remembers the elements for which
        the k-th bit is 0  */\\
/*  A1 remembers the elements for which
        the k-th bit is 1  */
\begin{verbatim}
    int n0;     //  size of A0
    int n1;     //  size of A1
    for (int k=0; k<t-1; k++) {
\end{verbatim}
/*  k –- number of the bit
        The numeration starts from 0.
        Does not check the sign bit  */
\begin{verbatim}
    n0=0;
    n1=0;
    for (int i=0; i<n; i++){
\end{verbatim}    	
/*  checks the k-th bit
        of the i-th element of the array: */
\begin{verbatim}
    if (A[i] & 1<<k) {
        A1[n1] = A[i];
        n1++;
        }
        else {
            A0[n0] = A[i];
            n0++;
            }
    }
\end{verbatim}
/*  We merge the two arrays.
        As a result we obtain an array whose
        elements if the k-th bit is equal to 0
        are at the beginning, and if it is equal to 1 – at the end. */
\begin{verbatim}
    for (int i=0; i<n0; i++)
        A[i] = A0[i];
    for (int i=0; i<n1; i++)
        A[n0+i] = A1[i];
    }
}
\end{verbatim}
/* The second function sorts the whole array */
\begin{verbatim}
template <class T>
void BitwiseSort(T A[],int n){
    T Aminus[n], Aplus[n];
\end{verbatim}
/* Aminus[n] -- An array with the negative values of A */\\
/* Aplus[n] -- An array with the nonnegative values of A */
\begin{verbatim}
    int Nm = 0, Np = 0;
\end{verbatim}
/* Nm -- number of elements written in Aminus */\\
/* Np -- number of elements written in Aplus */
\begin{verbatim}
    for (int i=0; i<n; i++) if (A[i] < 0) {
    Aminus[Nm] = A[i]; Nm++;
    }
        else {
        Aplus[Np] = A[i]; Np++;
        }
\end{verbatim}
/* Sorts the negative elements: */
\begin{verbatim}
	BitwiseSort1(Aminus,Nm);
\end{verbatim}
/* Sorts the nonnegative elements */
\begin{verbatim}
	BitwiseSort1(Aplus,Np);
\end{verbatim}
/*  We merge the two arrays:  */
\begin{verbatim}
    for (int i=0; i<Nm; i++)
        A[i] = Aminus[i];
    for (int i=Nm; i<n; i++)
        A[i] = Aplus[i-Nm];
}
\end{verbatim}

\section{Evaluation of the algorithm}

As a main disadvantage of the algorithm described by us comes the fact that it is applicable only to arrays of integers or symbols (type char). This is because for it we substantially use bitwise operations, which are applicable only over integer types of data. But this disadvantage is compensated by its high speed. As we will see below, algorithm \ref{biwisesortingalgor} works in time $O(n)$, where $n$ is the number of the elements which are subjected to sorting.

Except through the multiple experiments which we have made, with the help of the following theorem we will prove the correctness of the algorithm created by us.

\begin{theorem}
 During every execution of algorithm \ref{biwisesortingalgor} with an arbitrary input array of integers, as a result a sorted array is obtained.
\end{theorem}

Proof. It is enough to prove that function \verb"BitwiseSort1" works so as to fulfill the conditions of the theorem.

Let $A=\{ a_{0} ,a_{1} ,\ldots ,a_{n-1} \} $ be an arbitrary integer array with length $n$ and let $A^{(k)} =\{ a_{o}^{(k)} ,a_{1}^{(k)} ,a_{n-1}^{(k)} \} $ be the array which is obtained after iteration with number $k$, where $k=0,1,\ldots ,t-2$, $t$=sizeof(T)*8, i.e.\textit{ }$t$ is equal to the number of the bits which every element of $A$ occupies in the memory of the computer.

Let $x$ be an integer. For every natural number $k=0,1,2,\ldots $ we define the functions:

$$\mu_{k} (x)=x\; \% \; 2^{k} ,$$
where just like in programming languages C/C++ and Java the operator \% denotes the remainder during integer division. Apparently $\mu _{s-1} (x)=x$ if the absolute value of the integer $x$ can be written with no more than $s$ digits 0 or 1 in a binary notation. Therefore in order to prove that as a result of the work of the algorithm the array $A^{(t-2)} $ is sorted, it is enough to prove that the array $\overline{A^{(t-2)} }=\{ \mu _{t-2} (a_{0}^{(t-2)} ),\mu _{t-2} (a_{1}^{(t-2)} ),\ldots ,\mu _{t-2} (a_{n-1}^{(t-2)} )\} $ is sorted. Applying inductive reasoning, we will prove that for every $s$, such that $0\le s<t-1$, the array $\overline{A^{(s)} }=\{ \mu _{s} (a_{0}^{(s)} ),\mu _{s} (a_{1}^{(s)} ),\ldots \mu _{s} (a_{n-1}^{(s)} )\} $ is sorted.

When $s=0$ the assertion follows from the fact that during iteration with number 0 ($k=0$), $A^{(0)}$ is ordered so that first come all elements of the array which in their binary notation end in 0, followed by all elements of the array which in their binary notation end in 1.

We assume that for a certain natural number $s$, $0\le s<t-2$ the array $\overline{A^{(s)} }=\{ \mu _{s} (a_{0}^{(s)} ),\mu _{s} (a_{1}^{(s)} ),\ldots \mu _{s} (a_{n-1}^{(s)} )\} $ is sorted. But then analysing the work of the algorithm in $(s+1)$-th iteration, it is easy to see that the array A0, which is obtained from $\overline{A^{(s)} }$ taking in the same row only these elements of $A^{(s)} $ having 0 in bit with number $s+1$ , is a sorted array. Analogously we prove that the array A1 is sorted and in bit with number $s+1$ on each of its elements stands 1. Then the array $\overline{A^{(s+1)} }=\{ \mu _{s+1} (a_{0}^{(s+1)} ),\mu _{s+1} (a_{1}^{(s+1)} ),\ldots \mu _{s+1} (a_{n-1}^{(s+1)} )\} $, which is obtained from the merger of the arrays A0 and A1 where the elements of A0 precede the elements of A1, is sorted. And with this we have proven the theorem.

\begin{theorem}
Algorithm \ref{biwisesortingalgor} described with the help of programming language C++ works in time $O(n)$.
\end{theorem}

Proof. The assertion of the theorem follows from the fact that in function \verb"BitwiseSort1" we have only two nested loops. In the inner loop exactly $n$ iterations are performed, and in every iteration once the operation \verb"&" (bitwise conjunction), once the operation \verb"<<" (bitwise shift left), once the \verb"if" statement, once the assignment statement and once the increment statement  are performed. Each of the aforesaid operations is performed in constant time. The outer loop does $t-2$ iterations, where $t$ is a constant, and in every iteration besides the inner loops there are also two assignment operations.

In function \verb"BitwiseSort" the division of the array into two disjoing subarrays is performed apparently in time $O(n)$. The newly obtained two arrays are sorted in total time $O(n)$. The following merger of the two sorted arrays with total length $n$ is apparently also performed in time $O(n)$.


%

\end{document}